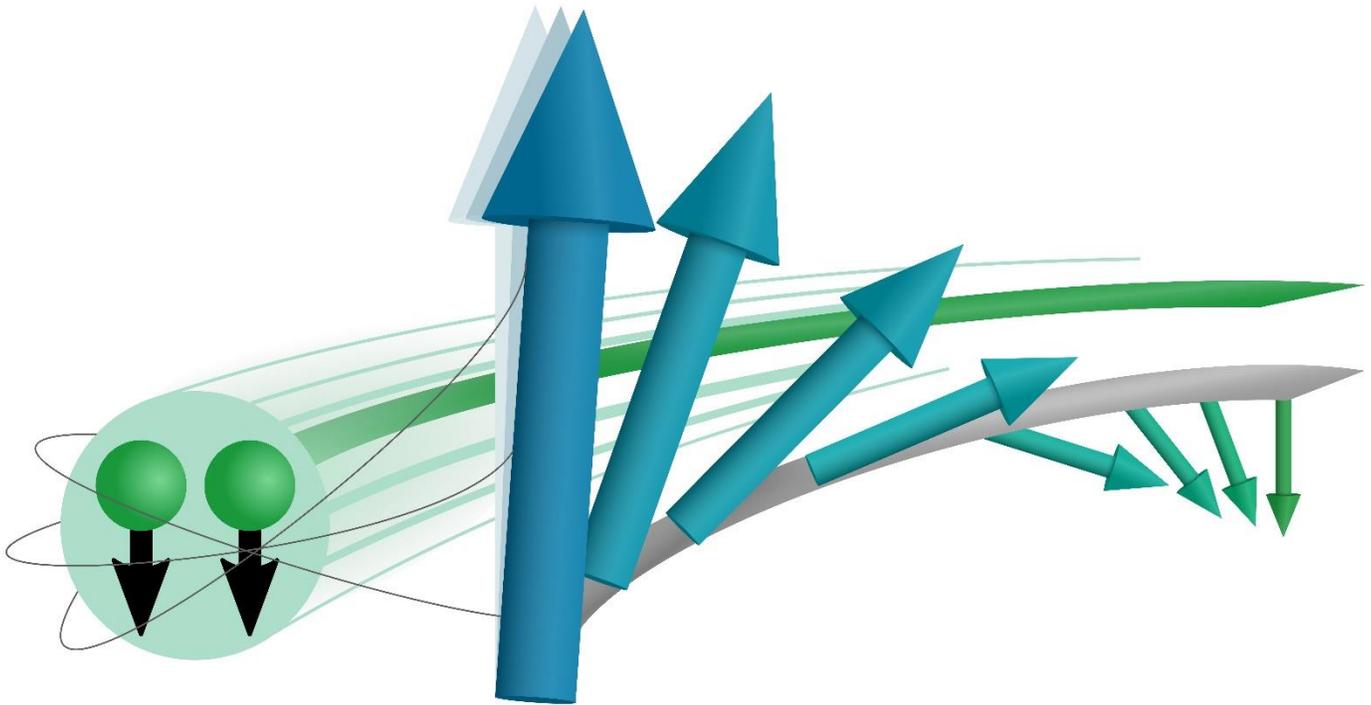

# USING SUPERCONDUCTIVITY TO CONTROL MAGNETISM:
# A FACET OF SUPERCONDUCTING SPINTRONICS

**Lina Johnsen Kamra and Akashdeep Kamra**

Condensed Matter Physics Center (IFIMAC) and Departamento de Física Teórica de la Materia Condensada, Universidad Autónoma de Madrid, E-28049 Madrid, Spain

Magnets are used in electronics to store and read information. A magnetic moment is rotated to a desired direction, so that information can later be retrieved by reading this orientation. Controlling the moment via electric currents causes resistive losses and heating, a major bottleneck in advancing computing technologies. Superconducting spintronics can resolve this using the unique features of superconductors.

## Manipulating a magnetic moment

Imagine a narrow hiking path. A devilish hiker has randomly turned all the signs along the path. Now, you need to switch them back before anyone gets lost. It is a straightforward task, but the path is long and full of obstacles. Now consider instead a different opportunity. The signs that you need to orient are on a road running along a river. The road and river are separated by a long pit with you on the river side. You can use a boat to easily ride down the river and along the road, but then it is hard to reach and correct the signs across the pit.

You need to find a new way, e.g., a stick, that can stretch across the pit and allow you to manipulate the signs while traveling with little effort on the boat.

The hiking path here is an analogue to a normal metal, in which an electric current attempting to orient a magnetic moment (our metaphorical signs) experiences scattering from obstacles and thus resistive loss. The road and river represent a superconductor, in which Cooper pairs can travel without resistance at the cost of seemingly not being able to influence the magnetic moments (Box 1). The goal is to utilize the unique properties of superconductors to efficiently manipulate magnetic moments, i.e., to find our metaphorical stick that enables orienting the signs while moving easily through the river. Below, we describe three examples where such means of using superconductors have been accomplished, closing our discussion by mentioning some opportunities and challenges for the future.

> **Box 1: Quasiparticles and Cooper pairs**
>
> The density of states in a superconductor as a function of the energy $E$ [panel (a)] bears a gap, within which there are no available electrons or "quasiparticles". Below the gap ($E < -\Delta$), all states are occupied. Above the gap ($E > \Delta$), the states are empty. When applying a voltage, the states above the gap can get occupied and carry electric currents with resistive losses. These currents are carried by quasiparticles – electron-like particles formed by a superposition of electrons and holes [panel (b)]. Within the gap, we have Cooper pair [panel (c)] states at zero energy. Consequently, these pairs can carry "supercurrents" without an energy cost, and thus, losses or heating.
>
> 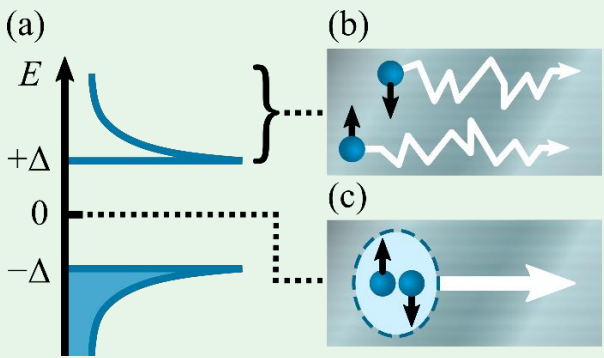

## Thermally induced domain wall motion

An important paradigm to store and manipulate information is using spatially varying magnetization, such as a domain wall[1] (Box 2). We can move a domain wall by sending a current of electrons through it (Box 2). Superconductors have no available electrons to carry currents at low energies due to the superconducting gap (Box 1). However, compensating for the lack of states in the gap, superconductors have a very large density of states at the gap edges. Furthermore, when we place the superconductor in contact with a ferromagnet, we need less energy to excite quasiparticles with spin along the ferromagnetic spin. When heating up one end of the superconductor, spin-polarized quasiparticles are excited and move from the hot end towards the cold one. Due to the high density of states at the gap edges in a superconductor, this thermoelectric effect is large[2] and can efficiently move domain walls[3], see Fig. 1(a).

## Changing the magnetic anisotropy

A magnet typically bears anisotropy resulting in a direction along which the magnetic moment prefers to align. This magnetic anisotropy is affected by the shape of the magnet such that the spins prefer to orient along the surface and not stick out. The magnetic moment or spins can also be locked to a special axis in the crystal. The preferred direction determines the magnetic ground state corresponding to the lowest energy.

At zero temperature, the free energy difference between the normal and superconducting states of a superconductor varies as $\sim |\Delta|^2$, where $2\Delta$ is the size of the superconducting gap. When placing a superconductor in contact with a magnet, the magnetic moment influences the gap and free energy in the former, and in this manner, the magnetic anisotropy of the total system. If we for instance sandwich a superconductor between two ferromagnets, the free energy associated with the superconducting condensate is minimized when the ferromagnets are antiparallel because the field from the magnets cannot suppress $\Delta$ as much[4]. To further control the magnetic anisotropy, we can employ a thin heavy metal layer between the superconductor and ferromagnet, thus introducing interfacial spin-orbit coupling. While in the normal state the magnetic spins prefer to align with the cubic crystal axes, they can be trapped at an in-plane 45°

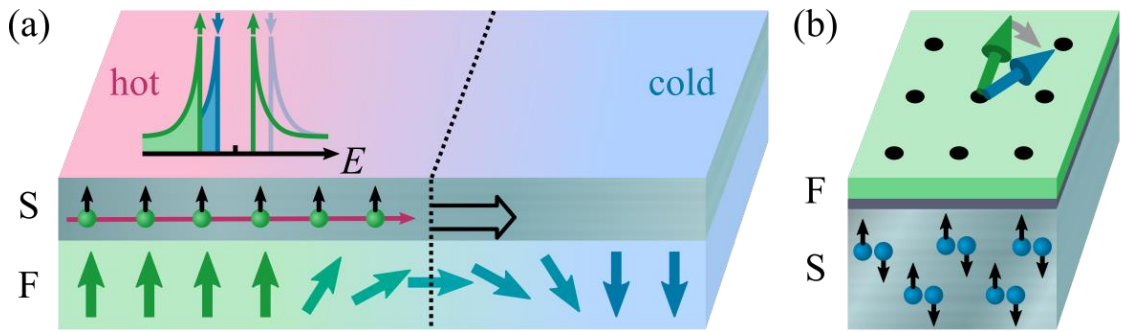

**Figure 1:** (a) When a superconductor (S) is in contact with a ferromagnetic domain, the density of states of the two spins (green and blue curves) are shifted so that the quasiparticles are mostly polarized along the spin of the ferromagnet (F). A heat gradient across the superconductor is therefore accompanied by a spin current (red arrow) of quasiparticles (green with black arrows) that can drive a domain wall (black dotted line)[4]. (b) When we decrease the temperature, the magnetic moment of a magnet can favor a different orientation (blue arrow) to minimize the free energy by allowing more Cooper pairs (blue with black arrows) to form in an adjacent superconductor. Here, the magnetic moment is rotated with respect to the crystal lattice (black dots) in the presence of interfacial spin-orbit coupling[7].

angle when lowering the temperature below the superconducting transition[5], see Fig. 1(b).

## Cooper pair currents for spin manipulation

While electrons tend to orient their spin along an adjacent ferromagnet's magnetization, Cooper pairs in a conventional superconductor have no net spin as the two electrons in the pair have opposite spins. When interfacing the superconductor to a ferromagnet, the spin of the Cooper pair along the magnetization remains conserved and zero. Its component perpendicular to this axis is however not conserved, as can be visualized by the up and down spins precessing in opposite directions around the axis. To engineer Cooper pairs that carry spin, we can therefore introduce noncollinear magnetic moments e.g., through misaligned ferromagnetic layers, or magnetic domain walls[6].

A challenge in using nondissipative Cooper pair currents for spin manipulation lies in the energy needed to alter a magnetic configuration. A nondissipative supercurrent reorienting a single magnet or driving a magnetic domain wall must be accompanied by dissipation somewhere[7]. Using our above analogy, if you use a stick to alter the signs while riding the boat down the river, the resistance from moving the signs will cause the boat to slow down. To keep up the speed, you need to row. This is still advantageous, since the resistance is only associated with orienting the signs, or spins, and not the transport itself.

## Beyond rotating static spins

Thus, the unique superconducting properties offer new and efficient means to manipulate magnetic configurations. The above discussion has focused on controlling static spins that point in a certain direction

> **Box 2: Noncollinearity and spin torques**
>
> 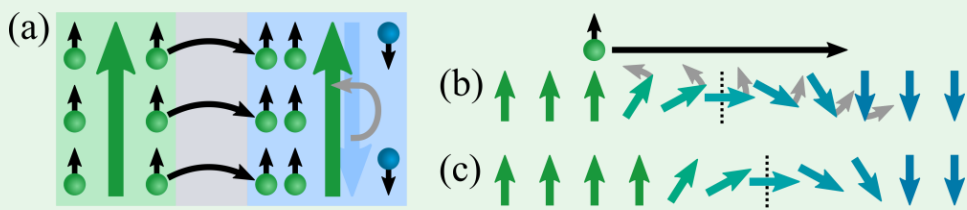
>
> Conduction electrons in a ferromagnet tend to align their spin with the magnetic moment. Driving a current can send electrons from one ferromagnet into another with opposite moment [panel (a)]. When they are in large enough numbers, they exert a spin torque in an attempt to align the magnetic moment with their spin, eventually causing the second ferromagnetic moment to flip (right green arrow) [panel (a)]. Similarly, a current of electrons through a domain wall will rotate the noncollinear magnetic moments [panel (b)] causing the domain wall (dotted line) to move [panel (c)].

before we try to rotate them. The use of supercurrents, however, can go well beyond. The dynamically moving magnetic moments of two ferromagnets can communicate with each other via spin supercurrents through an intermediate superconducting layer[8]. This allows for control of magnetization dynamics. Spins in magnets can moreover fluctuate giving rise to pure spin currents, or magnon currents. These can transport spin signals without a physical movement of electrons. Such magnons have recently been predicted to drift along with a supercurrent of spinful Cooper pairs in an unconventional superconductor[9]. This can result in a diode-like behavior when observing a magnon current along or opposite to the drift direction[10]. Magnons have also been predicted recently to induce and steer spinful Cooper pairs in a conventional superconductor, allowing supercurrents to mediate spin signals between two magnets[11]. The influence of superconductors on magnetic spins, magnetization dynamics, and magnon currents holds a high potential for intriguing scientific discoveries and addressing technological challenges facing mankind.

## About the authors

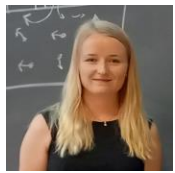

**Lina Johnsen Kamra** has obtained her PhD at the Norwegian University of Science and Technology (NTNU). Her PhD included stays at RIKEN in Japan and Universidad Autónoma de Madrid (UAM) in Spain. She is currently a postdoc at UAM.

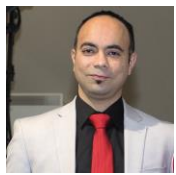

**Akashdeep Kamra** is a group leader and Ramón y Cajal fellow at UAM. After obtaining his PhD from the Delft University of Technology, he worked as a Humboldt research fellow at the University of Konstanz and as a senior researcher at NTNU.